\documentclass[aps,prd,a4paper,preprintnumbers,twocolumn,floatfix,showkeys,showpacs,nofootinbib]{revtex4}

\usepackage{amssymb}
\usepackage{subfigure}
\usepackage{color}
\usepackage{mathrsfs}
\usepackage{graphicx}
\def\be{\begin{equation}}
\def\ee{\end{equation}}
\def\bea{\begin{eqnarray}}
\def\eea{\end{eqnarray}}

\begin{document}

\title{Horizon thermodynamics and spacetime mappings}

\author{Valerio Faraoni}
\email[]{vfaraoni@ubishops.ca}
\affiliation{Physics Department and {\em STAR} Research Cluster, Bishop's University, Sherbrooke, Qu\'ebec, Canada J1M~1Z7
}

\author{Vincenzo Vitagliano}
\email[]{vincenzo.vitagliano@ist.utl.pt}
\affiliation{CENTRA, Instituto Superior T\'ecnico,\\ Universidade T\'ecnica de Lisboa - UTL, Av. Rovisco Pais 1, 1049 Lisboa, Portugal
}

\begin{abstract} When black holes are dynamical, event horizons are 
replaced by apparent and trapping horizons. Conformal and Kerr-Schild 
transformations are widely used in relation with dynamical black holes 
and we study the behaviour under such transformations of quantities 
related to the thermodynamics of these horizons, such as the 
Misner-Sharp-Hernandez mass (internal energy), the Kodama vector, 
surface gravity, and temperature. The transformation properties are not 
those expected on the basis of naive arguments.

\end{abstract}



\maketitle

\section{\label{section1}Introduction}

Black holes are one of the most intriguing predictions of General 
Relativity, which persist in alternative theories of gravity motivated 
by quantum considerations and by cosmology. Black holes have been shown 
to play very important roles in the dynamics and evolution of 
astrophysical systems and to be an important source of gravitational 
waves, which are expected to be detected experimentally within the 
next decade. 
In this regard, the prediction of accurate templates is a 
necessary condition for recovering a gravitational wave signal from the 
background noise. Waves generated during black hole collapse and the 
inspiral and merger of a binary system containing at least one 
black hole are notoriously 
difficult to predict accurately and the most sophisticated tools of 
approximate analytical and numerical relativity are used to 
attack the problems of the dynamics and wave generation in these 
systems. In numerical relativity, in particular, the ``black hole 
horizon'' is located using marginally trapped surfaces and apparent and 
trapping horizons ({\em e.g.}, \cite{numerical}): the teleological event 
horizon requires the knowledge of the entire future of the spacetime, which 
is impossible in a numerical calculation. On the contrary, the theories 
of black hole dynamics and thermodynamics constructed during the 1960s 
and 1970s \cite{Wald, WaldThermo, Poisson} rely on the concept of 
event horizon, which is well suited to static and stationary black 
holes. When black holes are dynamical, the concept of event horizon 
defining the black hole itself appears to be useless for practical 
purposes. It has been suggested that apparent horizons replace event 
horizons even though apparent horizons have two drawbacks: first, they 
depend on the spacetime foliation \cite{WaldIyer, SchnetterKrishnan} 
and, second, in some of the analytical solutions known to exhibit 
time-dependent apparent horizons, these can be timelike instead of 
spacelike.

It is claimed that it is the thermodynamics of apparent, not of event, 
 horizons which are physically meaningful 
\cite{Hayward1stlaw, AHthermodynamics}. 
However, while for stationary black holes several 
calculational methods give the same 
results for the surface gravity, the Hawking temperature and other 
thermodynamical quantities, this is not the case for dynamical black 
holes and for apparent/trapping horizons. There are several 
possibilities to define the surface gravity and horizon temperature and 
it is not clear which alternative is the physical one (see for example 
the reviews \cite{Pielhanetal, Nielsen2007}). The issue is not an easy 
one to settle and the debate will likely remain for some time. At the 
same time, the thermodynamics of apparent horizons are being criticized 
(for example, apparent horizons may not satisfy a quantum-generalized 
second law \cite{Wall}), but apparent horizons and marginally trapped 
surfaces are all that is used to locate the black hole ``horizon'' in 
numerical relativity \cite{numerical}. Here we do not argue in favour of 
one prescription or the other but we focus on providing tools to 
understand the various proposals for thermodynamical quantities 
(internal energy \cite{MisnerSharp, MisnerHernandez, Hawkingquasilocal, 
Haywardquasilocal},
 Kodama vector, Kodama surface gravity \cite{Kodama}, and temperature).

It is useful to have explicit examples of dynamical black holes to use as 
testbeds for alternative definitions and for new approaches. While 
analytical solutions describing dynamical black holes are rare, a few are 
known which are interpreted as black holes embedded in cosmological 
backgrounds, in General Relativity and in alternative theories of gravity 
(see Ref.~\cite{mydynBHreview} for a recent review). These are all 
spherically symmetric solutions and, because of their relative simplicity, 
we also focus on spherical symmetry in this paper. Another important reason is 
that, in General Relativity, the Misner-Sharp-Hernandez mass is defined 
only in spherical symmetry. An added advantage is that in spherical 
symmetry one has preferred spacetime slicings and the problem of the 
foliation-dependence of the apparent horizons is alleviated.  We discuss 
two quantities appearing in black hole thermodynamics, the 
Misner-Sharp-Hernandez mass (which coincides with the Hawking-Hayward 
quasi-local energy in spherical symmetry and is identified with the 
internal energy $U$ in the first law $TdS=dU+dW$) 
and the Kodama vector (which defines a 
Kodama surface gravity and horizon temperature entering the first law of 
thermodynamics).

Within General Relativity, and even more in scalar-tensor and $f(R)$ 
gravity, conformal transformations are a useful mathematical tool to relate 
analytical solutions. It is not surprising, therefore, that following 
earlier literature on the interplay between conformal transformations and 
horizon thermodynamics in asymptotically flat spacetimes in various 
theories of gravity \cite{JacobsonKang, KogaMaeda, JacobsonKangMyers} and 
for conformal Killing horizons \cite{confoKilling}, recent papers have 
discussed the thermodynamics of apparent horizons in relation with 
conformal transformations \cite{NielsenVisser06, DeruelleSasaki, NielsenCT, 
AbreuVisser10, ValerioAlex, MarquesRodrigues12, AlexFirouzjaee}. Another 
issue is the quantum state of the scalar field used to compute the 
Hawking effect \cite{VisserBarceloLiberatiSonego}: in general the vacuum state 
 will be changed by a conformal transformation.

The first and most obvious property of conformal transformations in this 
context is that the location of apparent horizons changes under conformal 
transformations or apparent horizons are created where there was 
none.\footnote{For example, the Husain-Martinz-Nu\~nez scalar field 
solution of General Relativity \cite{HusainMartinezNunez}, 
which hosts a black hole in a portion of the 
 spacetime manifold, is conformal to the 
Fisher-Janis-Newman-Winicour-Wyman solution which describes a naked 
singularity \cite{Fisher}.} On the contrary event horizons are null 
surfaces and are unaffected by conformal rescalings. A precise way to 
recover apparent horizons under conformal mappings was discussed in 
Ref.~\cite{ValerioAlex}: a surface of zero entropy expansion should be 
considered instead of a trapping horizon with vanishing area expansion. 
This procedure preserves the thermodynamical properties of quasi-local 
horizons under conformal transformations \cite{ValerioAlex}. This 
non-invariance of the apparent horizons should be kept in mind in the 
following discussion in Sec.~\ref{section2}. For dynamical horizons, the 
Kodama vector is often used in place of a timelike Killing vector to 
identify a preferred notion of surface gravity and horizon temperature. 
The transformation of these quantities under conformal transformations 
is discussed in Sec.~\ref{section3}.

Another technique which is used to generate cosmological black hole 
solutions starting from stationary black hole metrics as ``seeds'' is 
that of the Kerr-Schild transformation. The transformation properties of 
the relevant thermodynamical quantities under Kerr-Schild 
transformations is discussed in Sec.~\ref{K-S}, while 
Sec.~\ref{Discussion} contains our conclusions.

\section{\label{section2}Behavior of the Misner-Sharp-Hernandez mass under 
conformal transformations}

The Misner-Sharp-Hernandez mass $M_{MSH}$ 
\cite{MisnerSharp, MisnerHernandez}, 
which coincides with the 
Hawking-Hayward quasi-local mass \cite{Hawkingquasilocal, Haywardquasilocal} 
in spherical symmetry, is usually 
identified with the internal energy $U$ of a black hole 
system, which enters the first 
law of thermodynamics $TdS=dU+dW$ in General Relativity with spherical 
symmetry. Here we derive a general law for the transformation of the 
Misner-Sharp-Hernandez mass under conformal transformations of the 
metric $g_{ab}\rightarrow \tilde{g}_{ab}=\Omega^2 g_{ab}$. We 
consider a general spherically symmetric line element of the 
form 
\be \label{3bis}
ds^2 = -A(t,R)dt^2 +B(t,R)dR^2 +R^2 d\Omega_{(2)}^2 \,, 
\ee 
where $R$ is the areal radius determined by the 2-spheres of symmetry 
and $d\Omega_{(2)}^2=d\theta^2 +\sin^2 \theta d\varphi^2$ is the line 
element on the unit 2-sphere. The conformal transformation is assumed to 
preserve the spherical symmetry, {\em i.e.}, $\Omega=\Omega(t, R)$. Note 
that, in general, if the metric $g_{ab}$ is a solution of the Einstein 
equations with a ``reasonable'' form of matter, $\tilde{g}_{ab}$ is not 
a solution for the same form of matter, or the corresponding form of 
mass-energy can be completely unrealistic. A vacuum solution 
$g_{ab}$ will be mapped into a non-vacuum solution. However, the use of 
conformal transformations  is so widespread even in GR to motivate 
the 
following analysis.

The line element (\ref{3bis}) is mapped into
\be\label{3ter}
d\tilde{s}^2 = \Omega^2 ds^2=-\Omega^2 Adt^2 +\Omega^2BdR^2 
+\tilde{R}^2 d\Omega_{(2)}^2 \,,
\ee
where 
\be \label{4}
\tilde{R}=\Omega R
\ee
is the ``new'' areal radius. It is well known since the early work of 
Dicke \cite{Dicke} that, under a conformal transformation, lengths and 
time intervals scale as $\Omega$ (which is consistent with eq.~(\ref{4})) 
while masses and energies scale as 
$\Omega^{-1}$.  Therefore, one would naively expect the 
Misner-Sharp-Hernandez mass to scale as $\Omega^{-1}$ and the Hawking 
temperature to scale the same way since $k_BT$ (where $k_B$ is the 
Boltzmann constant) is dimensionally an energy. The situation is, 
however, more complicated: Dicke's dimensional argument \cite{Dicke} 
applies to test particles but the Misner-Sharp-Hernandez/Hawking-Hayward 
mass includes gravitational energy and other energy contributions, and 
the Hawking temperature results from a process involving a quantum test 
field, hence classical test particle considerations may not be 
sufficient (indeed, we will find that this is the case, at least for 
the mass $M_{MSH}$).

To find the transformation properties of the Misner-Sharp-Hernandez mass 
$M_{MSH}$ of a sphere of radius $R$, consider its 
definition in the tilded and in the non-tilded worlds, 
\be \label{5} 
1-\frac{2\tilde{M}_{MSH}}{ \tilde{R}} = \tilde{g}^{ab}
\tilde{\nabla}_a \tilde{R} \tilde{\nabla}_b \tilde{R}
\ee
and 
\be \label{6} 
1-\frac{2M_{MSH}}{R} = g^{ab}
\nabla_a R  \nabla_b R \,.
\ee
Using eq.~(\ref{4}), eq.~(\ref{5}) gives 
\be 
\Omega^{-2} g^{ab} \nabla_a \left( \Omega R \right)
\nabla_b \left( \Omega R \right) = 1-\frac{2 \tilde{M}_{MSH}}{\Omega R} \,,
\ee
from which it follows that 
\be\label{7}
\tilde{M}_{MSH}=\Omega M_{MSH} -\frac{R^3}{2\Omega}\, 
\nabla^a \Omega\nabla_a \Omega 
-R^2 \nabla^a \Omega  \nabla_a R \,.
\ee

The conformal transformation turns geometry ($\Omega$ and its gradient) 
into quasi-local mass-energy, as is evident from the fact that a vacuum 
solution is turned into a non-vacuum one due to the transformation 
property of the Ricci tensor \cite{Synge, Wald, 
mybook}  
\begin{eqnarray}
\tilde{R}_{ab} & = & R_{ab}- 2\nabla_a\nabla_b ( \ln \Omega) -g_{ab} 
g^{ef}\nabla_e \nabla_f (\ln \Omega) \nonumber\\
&&\nonumber\\
&\, & + 2 \nabla_a (\ln \Omega) \nabla_b (\ln \Omega) 
-2 g_{ab} g^{ef} \nabla_e (\ln \Omega) \nabla_f (\ln \Omega) 
\nonumber\\
\end{eqnarray}
(the ambiguity in the separation of matter and geometry is a 
recurrent feature of 
alternative theories of gravity \cite{ThomasStefanoValerioIJMPD}). 
Definitely, $M_{MSH}$ does not scale as $\Omega^{-1}$, as could naively be 
expected from Dicke's dimensional argument \cite{Dicke}.

\subsection{Example: FLRW space}

Let us look at a very simple example, the spatially flat 
Friedmann-Lema\^{i}tre-Robertson-Walker space with line element 
\be
d\tilde{s}^2=-dt^2 +a^2(t) d\vec{x}^2 \,.
\ee

By using the conformal time $\eta$ related to the comoving time $t$ by 
$dt=ad\eta$, the line element is manifestly conformal to the Minkowski 
one,
\be
d\tilde{s}^2=a^2(\eta) \left( -d\eta^2 +dr^2+r^2 d\Omega_{(2)}^2 
\right)=\Omega^2 ds^2 \,,
\ee
where we have used spherical coordinates $\left( \eta, r, \theta, 
\varphi \right)$ for the Minkowski line element $ds^2$ and the conformal 
factor $\Omega=a(\eta)$ preserves the spherical symmetry about every 
spatial point. A sphere of radius $r$ (areal radius) in Minkowski space 
has vanishing Misner-Sharp-Hernandez/Hawking-Hayward mass, $M_{MSH}=0$, 
while the mass of the corresponding sphere in FLRW space is, according to 
eq.~(\ref{7}),
\be\label{boh}
\tilde{M}_{MSH}=a M_{MSH}-\frac{r^3}{2a} \left( -a_{, \eta}^2 \right) 
-r^2 \nabla^c a(\eta)\nabla_c r=\frac{r^3}{2a} \, a_{,\eta}^2 \,.
\ee
Since  $a_{,\eta}=a\, da/dt\equiv a\dot{a} $, and the areal 
radius of FLRW space is $\tilde{R}=ar$, one obtains from eq.~(\ref{boh}) that 
\be \label{8} 
\tilde{M}_{MSH}=\frac{4\pi}{3} \tilde{R}^3 \rho \,, 
\ee 
the well known expression of the Hawking-Hayward quasi-local mass in 
this space \cite{Haywardquasilocal}, corresponding to the mass of cosmic 
fluid enclosed in a sphere of proper radius $\tilde{R}$.

\subsection{Example: the Sultana-Dyer cosmological black hole}

The Sultana-Dyer spacetime \cite{SultanaDyer} is a an inhomogeneous and 
time-dependent solution of GR interpreted as a black hole embedded in a 
spatially flat FLRW ``background'' (the quotation marks are compulsory since one cannot 
covariantly split a metric into a ``background'' and a ``deviation'' 
from it due to the non-linearity of the Einstein equations). The matter 
source is composed of two non-interacting perfect fluids, a timelike 
dust and a null dust \cite{SultanaDyer}. 
The energy density becomes negative at late times 
near an event horizon. This solution has been studied as an example of a 
time-dependent black hole horizon for which the Hawking temperature can 
be derived explicitly \cite{SaidaHaradaMaeda} to shed light on the 
Hawking effect and the thermodynamics of time-evolving horizons. In this 
context, it is not important that the metric arises as a 
solution of GR with ill-behaved matter, or even that the metric is a 
solution of GR or another theory of gravity.  The line element of 
the Sultana-Dyer metric is usually written as
\be 
d{s}^2 = a^2(\tau) \left[ - d\tau^2 
+{dr^2}+r^2 d\Omega_{(2)}^2+\frac{2m}{r}(d\tau+dr)^2 \right] \,.
\ee 
What is interesting here is that the Sultana-Dyer 
metric can be obtained by conformally transforming the Schwarzschild 
solution. The line element can be rewritten as \cite{SultanaDyer, SaidaHaradaMaeda} 
\be \label{8plus1}
d\tilde{s}^2 = a^2(\tau) \left[ -\left(1-\frac{2m}{r}\right) d\eta^2 
+\frac{dr^2}{1-\frac{2m}{r}} +r^2 d\Omega_{(2)}^2 \right] \,, 
\ee 
where 
$a(\tau)=\tau^2 $ is the scale factor of the spatially flat 
FLRW `background'' (to 
which the solution reduces when $m\rightarrow 0$ or for $r\rightarrow 
+\infty$) and \cite{SultanaDyer, SaidaHaradaMaeda} 
\be \label{8plus3}
\tau\left( \eta, r \right)= \eta+2m \ln \left( 
\frac{r}{2m}-1 \right) \,. 
\ee 
The comoving time $t$ of the FLRW ``background'' is given by 
$dt=ad\eta$, where $\eta$ is the conformal time. This example 
contains the previous FLRW example as the trivial case $m=0$. 
The FLRW ``background'' has scale factor $a(t)\sim t^{2/3}$ (the time 
evolution caused by a dust) in terms of comoving time. The 
Misner-Sharp-Hernandez mass of a sphere of radius $r$ 
is easily computed from the line element 
(\ref{8plus1}) using the definition (\ref{5}) \cite{SaidaHaradaMaeda}:
\be\label{8plus4}
\tilde{M}_{MSH}=ma-2mra_{,\tau} +\frac{r^3a_{,\tau}^2}{2a} 
\left(1+\frac{2m}{r} \right)\,.
\ee
Since the Sultana-Dyer metric is conformal to the Schwarzschild one 
which has $M_{MSH}=m$ and $R=r$ with $\Omega=a$, we can apply 
eq.~(\ref{7}) which gives, using eq.~(\ref{8plus3})
\begin{eqnarray}
\tilde{M}_{MSH} &=& ma-\frac{r^3}{2a} \left[ g^{00}\left( a_{,\tau} 
\tau_{, \eta} \right)^2 +g^{11} \left( a_{, \tau} \tau_{, r} 
\right)^2 \right] \nonumber\\
&&\nonumber\\
&\, & 
-r^2g^{11} a_{,\tau}\tau_{,r} \nonumber\\ 
&&\nonumber\\ 
&=& ma + \frac{ r^3 a_{,\tau}^2}{2a} 
\left( 1+\frac{2m}{r} \right) -2mra_{,\tau} \,. \label{8plus9}
\end{eqnarray}
Hence, also for this example, eq.~(\ref{7}) gives the correct result 
(\ref{8plus4}) computed independently in Ref.~\cite{SaidaHaradaMaeda}. 
The Misner-Sharp-Hernandez mass does not scale as $\Omega^{-1}$, rather 
it has a contribution which scales as $\Omega$ (comoving with the cosmic 
substratum) plus two contributions of opposite signs which depend on the 
expansion rate and and on the expansion rate squared, respectively.

The interpretation of the result (\ref{8plus9}) (not attempted in 
Ref.~\cite{SaidaHaradaMaeda}) is not straightforward. One can use the 
identity $a_{,\eta}=a_{, \tau}= a \, \partial a/\partial t \equiv 
a\dot{a}$, the 
definition of the (comoving) Hubble parameter $H\equiv \dot{a}/a$, and 
straightforward algebra to write
\be \label{8plus10}
\tilde{M}_{MSH}= 
m \, a\left( 1-H\tilde{R}\right)^2 +\frac{H^2\tilde{R}^3}{2} \,, 
\ee 
where $\tilde{R}=ar$ is the areal radius of the Sultana-Dyer spacetime. 
According to eq.~(\ref{8plus10}), the mass $\tilde{M}_{MSH}$ consists of 
two contributions. The first contribution is the mass $m=M_{MSH}$ of the 
Schwarzschild ``seed'' metric scaled by the conformal factor $a$ but 
diluted by the expansion of the universe by the factor 
$\left(1-H\tilde{R}\right)^2$ (this factor vanishes at 
$\tilde{R}=H^{-1}$, the radius that the cosmological horizon would have 
were the central inhomogeneity absent). It is not entirely clear how to 
interpret the competing effects of the factors $a$ and $\left( 
1-H\tilde{R}\right)^2$ in this contribution to $\tilde{M}_{MSH}$. The 
second contribution to $\tilde{M}_{MSH}$ can be written as 
$\frac{4\pi}{3} \, \tilde{R}^3 \rho$, where $\rho =\frac{3 H^2}{8\pi} $ 
is the density that the cosmological fluid would have in the absence of 
the central inhomogeneity, to which the real density reduces for values 
of the radial coordinate $r \gg m$. This second contribution is 
interpreted as the mass of the cosmological fluid contained in the 
sphere of areal radius $\tilde{R}$ (or coordinate radius $r$). In any 
case, the expression (\ref{8plus10}) makes it clear that the 
Misner-Sharp-Hernandez mass $\tilde{M}_{MSH}$ of the Sultana-Dyer 
spacetime is always positive if the mass $m$ of the seed Schwarzschild 
metric is positive (or zero, which corresponds to $g_{ab}$ being the 
Minkowski metric and $\tilde{g}_{ab}$ the spatially flat FLRW space).

\section{\label{section3}Behavior of the Kodama vector, surface gravity, and 
temperature under conformal transformations}

In the spacetime with spherically symmetric metric (\ref{3bis}) the 
Kodama vector is given by
\be\label{11}
K^a=\frac{1}{\sqrt{AB}}\, \left( \frac{\partial }{\partial t} \right)^a \,,
\ee
with a similar expression holding in the rescaled spherical metric 
(\ref{3ter}). In order to use the tilded version of this expression, 
the metric (\ref{3ter}) must 
be recast in terms of the areal radius $\tilde{R}$ in the form 
\be\label{11bis}
d\tilde{s}^2 = -\tilde{A}dT^2 +\tilde{B}d\tilde{R}^2 +\tilde{R}^2 
d\Omega_{(2)}^2 \,,
\ee
where $T$ will be, in general, a new time coordinate which is a function 
of both $t$ and $R$. Since $\Omega=\Omega\left( t, R\right) $ to 
preserve the spherical symmetry and $\tilde{R}=\Omega R$, the relation 
between differentials
\be\label{12}
dR=\frac{d\tilde{R} - \Omega_{,t}Rdt}{ \Omega_{,R}R+\Omega} 
\ee
follows. Inserting this relation in the line element (\ref{3ter}) yields
\begin{eqnarray}
d\tilde{s}^2 &=& -\left[ \Omega^2A 
-\frac{ \Omega_{,t}^2R^2\Omega^2 B}{ 
\left( \Omega_{,R}R+\Omega \right)^2}\right] dt^2 \nonumber\\
&&\nonumber\\ 
&\, & +\frac{ \Omega^2B}{ \left( \Omega_{,R}R+\Omega \right)^2 }\, 
d\tilde{R}^2 \nonumber\\
&&\nonumber\\
&\, & -\, \frac{2\Omega^2 \Omega_{,t}BR}{ 
\left( \Omega_{,R}R+\Omega \right)^2}\, dtd\tilde{R} 
+\tilde{R}^2 d\Omega_{(2)}^2 \,. \label{11ter}
\end{eqnarray} 
In order to achieve the diagonal form  of the line element, 
the $dt \, d\tilde{R}$ cross-term must be eliminated by introducing a new 
time coordinate $T(t, R)$ defined by
\be \label{13}
dT=\frac{1}{F} \left( dt +\beta d\tilde{R} \right) \,,
\ee
where $\beta\left( t, \tilde{R} \right) $ is a function to be determined 
and $F\left(t, \tilde{R} \right)$ is an integrating factor which must 
satisfy the equation
\be\label{eqforF}
\frac{\partial}{\partial \tilde{R}} \left( \frac{1}{F} \right) = 
\frac{\partial}{\partial t} \left( \frac{\beta}{F} \right) 
\ee 
to guarantee that $dT$ is an exact differential. The substitution of 
$dt=FdT-\beta d\tilde{R}$ into eq.~(\ref{11ter}) yields
\begin{widetext}
\begin{eqnarray}
d\tilde{s}^2 &=& -\left[ \Omega^2A -\frac{ \Omega_{,t}^2R^2\Omega^2 B}{ 
\left( \Omega_{,R}R+\Omega \right)^2}\right] F^2 dT^2  +
\left\{ -\beta^2 \left[ \Omega^2A -\frac{ \Omega_{,t}^2R^2\Omega^2 B}{ 
\left( \Omega_{,R}R+\Omega \right)^2}\right] 
+\frac{ \Omega^2 B}{ \left( \Omega_{,R}R+\Omega \right)^2 } \right.\nonumber\\
&&\nonumber\\
&\, & \left. 
+\frac{2 \beta \Omega_{,t}\Omega^2 BR}{ \left( \Omega_{,R}R+\Omega \right)^2}
\right\} d\tilde{R}^2 
 +2F \left\{ \beta \left[ \Omega^2A -\frac{ \Omega_{,t}^2R^2\Omega^2 
B}{ \left( \Omega_{,R}R+\Omega \right)^2}\right] - \frac{ 
\Omega_{,t}\Omega^2 R B}{ \left( \Omega_{,R}R+\Omega \right)^2} 
\right\} dT \, d\tilde{R} +\tilde{R}^2 d\Omega_{{2}}^2 \,. 
\end{eqnarray} 
The choice 
\be\label{16} 
\beta \left( t, R \right) = 
\frac{ \Omega_{,t} \Omega^2 BR}{ \left[ \Omega^2A -\frac{ 
\Omega_{,t}^2R^2\Omega^2 B}{ \left( \Omega_{,R}R+\Omega 
\right)^2}\right] \left( \Omega_{,R}R+\Omega \right)^2} 
\ee 
then turns the 
line element into the diagonal form 
 \be 
d\tilde{s}^2 = 
-\left[ \Omega^2A -\frac{ \Omega_{,t}^2\Omega^2 BR^2}{ \left( 
\Omega_{,R}R+\Omega \right)^2}\right] F^2 dT^2 
+ \frac{ \Omega^2 B}{ 
\left( \Omega_{,R}R+\Omega \right)^2 } 
\left\{ 1+ \frac{ \Omega_{,t}^2 \Omega^2 BR^2}{ 
\left( \Omega_{,R}R+\Omega \right)^2  \left[ \Omega^2 
A -\frac{ \Omega_{,t}^2\Omega^2 BR^2}{ \left( \Omega_{,R}R+\Omega 
\right)^2 } \right]} \right\} d\tilde{R}^2 +\tilde{R}^2 d\Omega_{{2}}^2 
\,.\label{17} 
\ee 
The comparison of this line element 
with eq.~(\ref{11bis}) yields 
\begin{eqnarray} 
\tilde{A} &=& \left[ 
\Omega^2A -\frac{ \Omega_{,t}^2\Omega^2 BR^2}{ \left( 
\Omega_{,R}R+\Omega \right)^2}\right] F^2 \,,\label{18}\\ &&\nonumber\\ 
\tilde{B} &=& \frac{B\Omega^2}{ \left( \Omega_{,R}R+\Omega \right)^2} 
\left\{ 1+ \frac{ \Omega_{,t}^2 \Omega^2 BR^2}{ \left( 
\Omega_{,R}R+\Omega \right)^2 \left[ \Omega^2 A -\frac{ 
\Omega_{,t}^2\Omega^2 BR^2}{ \left( \Omega_{,R}R+\Omega \right)^2 } 
\right]} \right\} \,. \label{19} 
\end{eqnarray} 
\end{widetext} 
In the non-tilded world (\ref{3bis}) 
the unit vector in the time direction is $ u^a 
\equiv \left( \frac{\partial}{\partial t} \right)^a $ and has components 
$
u^{\mu}=\left( \frac{1}{\sqrt{A}}, \vec{0} \right)$, while the 
corresponding vector $v^a \equiv \left( \frac{\partial}{\partial T} 
\right)^a $ in the tilded world (\ref{3ter}) has components $ v^{\mu}=\left( 
\frac{1}{\sqrt{\tilde{A}}}, \vec{0} \right) $.  The corresponding Kodama 
vectors have components $K^{\mu}=\left( \frac{1}{A\sqrt{B}}, \vec{0} 
\right)$ (in coordinates $\left( t, R, \theta, \varphi \right)$) and 
$\tilde{K}^{\mu}=\left( \frac{1}{\tilde{A}\sqrt{\tilde{B}}}, 
\vec{0} \right)$ (in 
coordinates $\left( T, \tilde{R}, \theta, \varphi \right)$). Therefore, 
the only non-vanishing component of the Kodama vector satisfies
\be\label{24}
\tilde{K}^0 = \frac{ A 
\left( \Omega_{,R}R+\Omega \right)}{
\tilde{A} \Omega \sqrt{
1+ \frac{ \Omega_{,t}^2 \Omega^2 BR^2}{ \left( 
\Omega_{,R}R+\Omega \right)^2 + \left[ \Omega^2 A -\frac{ 
\Omega_{,t}^2\Omega^2 BR^2}{ \left( \Omega_{,R}R+\Omega \right)^2 } 
\right] } } }\, K^0  \,.
\ee

The norm squared of the Kodama vector  in the non-tilded metric is 
\be
K^a K_a= -A (K^0)^2
\ee
and the norm squared of $\tilde{K}^a$ is 
\begin{eqnarray}
\tilde{K}^a\tilde{K}_a &=& -\frac{ A
\left( \Omega_{,R}R+\Omega \right)^2 (K^0)^2}{
\Omega^4  F^2} \nonumber\\
&&\nonumber\\
&=& \frac{ \left( \Omega_{,R}R+\Omega \right)^2}{
\Omega^4 F^2 } \, \left( K^a K_a \right) \,.
\label{30}
\end{eqnarray}

The factor $ \frac{ \left( \Omega_{,R}R+\Omega \right)^2}{ \Omega^4 F^2} 
$ is non-negative and therefore $ \tilde{K}^a $ has the same causal 
nature as $K^a$ except possibly where this factor vanishes, which is 
excluded by the following considerations. It would seem that 
eq.~(\ref{30}) singles out a class of special conformal transformations, 
those which satisfy identically the equation 
\be \label{31} 
\Omega_{,R}R+\Omega =0 \,, 
\ee 
but these transformations are physically irrelevant. In fact, 
eq.~(\ref{31}) is immediately integrated to
\be \Omega \left(t, R \right)=\frac{f(t)}{R} 
\ee 
with $f(t)$ a positive function of the time coordinate. In addition to 
being ill-defined at $R=0$ and $R=+\infty$, this transformation changes 
the causal character of the areal radius since eq.~(\ref{4}) then gives 
the areal radius $\tilde{R}=f(t)$ and $\tilde{R}$ becomes a 
timelike coordinate. We will not consider these conformal transformations 
further and we will assume that $ \Omega_{,R}R+\Omega$ does not vanish 
identically.

The Kodama surface gravity $\kappa_K$ is given as follows: let the 
spherically symmetric spacetime metric be given by 
\be 
ds^2 = h_{ij}dx^i dx^j+R^2  d\Omega_{(2)}^2 \,, 
\ee 
where $h_{ij}$ ($i,j=0,1$) is the 2-metric in the space orthogonal to the 
2-spheres of symmetry and $R$ is the areal radius; then
\be \label{surfacegravity}
\kappa_{K} =\frac{1}{2} \ \Box_{(h)} R
=\frac{1}{2\sqrt{-h}}\, \partial_i 
\left( \sqrt{-h} \, h^{ij} \partial_j R \right)
\ee
where $h$ is the determinant of $h_{ij}$. Since, in our notations, it is 
$h_{ij}=$diag$(-A, B)$, $h=-AB$, $\tilde{h}_{ij}=$diag$( -\Omega^2 A, 
\Omega^2 B )$, $\tilde{h}=\Omega^4 h$, and $\tilde{R}=\Omega R$, the 
Kodama surface gravity in the tilded world is 
\begin{eqnarray} 
\tilde{\kappa}_K &=& \frac{1}{2\sqrt{-\tilde{h} }}\, \partial_i \left( 
\sqrt{-\tilde{h}} \, \tilde{h}^{ij} \partial_j \tilde{R} \right) 
\nonumber\\ 
&&\nonumber\\ 
&=& \frac{1}{2\Omega^2 \sqrt{-h}}\, \partial_i 
\left( \sqrt{-h} \, h^{ij} \Omega \partial_j R + \sqrt{-h} \, h^{ij} R 
 \partial_j \Omega \right) \nonumber\\ 
&&\nonumber\\ 
&=& 
\frac{\kappa_K}{\Omega} + \frac{1}{2\Omega^2 \sqrt{-h}}\, \partial_i 
\left( \sqrt{-h} \, h^{ij} R \partial_j \Omega \right) \nonumber\\
&&\nonumber\\ 
&\,&  +\frac{1}{2\Omega^2} \, h^{ij} \partial_i \Omega \partial_j R 
\,. \label{Nexo}
\end{eqnarray} 
To the best of our knowledge, this transformation formula was first 
presented in the static case in Ref.~\cite{AlexFirouzjaee} (see their 
eq.~(29) and  note that the last term in eq.~(\ref{Nexo}) vanishes 
on the apparent horizon if $\Omega=\Omega(R)$ only). We can 
further write it as 
\begin{eqnarray} 
\tilde{\kappa}_K &=& \frac{\kappa_K}{\Omega} \nonumber\\
&&\nonumber\\
&\, & + \frac{1}{2\Omega^2 \sqrt{AB}} \left\{ -R \left[ \Omega_{,tt} 
\sqrt{\frac{B}{A}} +\frac{\Omega_{,t}\left( AB_{,t} -BA_{,t} 
\right)}{2A\sqrt{AB} } \right] \right.\nonumber\\
&&\nonumber\\
&\, &  +\sqrt{\frac{A}{B}} \, \Omega_{,R} 
+R\sqrt{\frac{A}{B}} \, \Omega_{,RR} \nonumber\\
&&\nonumber\\
&\, & \left. +\Omega_{,R}R \, \frac{\left( 
B\, A_{,R} -A \, B_{, R} \right)}{B\sqrt{AB} } \right\} 
+ \frac{\Omega_{, R}}{ 2\Omega^2 B}  \,. 
\end{eqnarray}
Therefore, the Kodama temperature 
\be
T_K=\frac{\kappa_K}{2\pi}
\ee 
(evaluated at the apparent horizon) does not scale simply as $ 
\tilde{T}_K \sim T_K/\Omega$, as it could be expected naively from 
Dicke's dimensional argument (\cite{Dicke}, see also \cite{myHawkingT}). 
However, a simplification is possible in the case of cosmological black 
holes obtained via a conformal transformation of a stationary black hole 
with conformal factor equal to the scale factor $a$ of a ``background'' 
FLRW space. In this case the conformal factor depends only on time and 
the transformation property of the Kodama temperature simplifies to
\be
\tilde{T}_K = \frac{T_K}{\Omega}-\frac{R}{4\pi \Omega^2}  
\left[ \frac{\Omega_{,tt}}{A}   
+\frac{\Omega_{,t}\left( AB_{,t} -BA_{,t} \right)}{2A^2 B} \right] 
 \,.\label{Ttilde} \nonumber\\
\ee
It seems intuitive that an apparent horizon evolving arbitrarily fast would not 
constitute a system in thermodynamical equilibrium and non-equilibrium 
thermodynamics would be necessary, as opposed to a stationary black hole 
which is instead a system in thermodynamical equilibrium. These 
would be bad 
news for the thermodynamics of apparent/trapping horizons; however, it is 
reasonable to expect that an adiabatic approximation in which the 
spacetime and its apparent horizons evolve slowly constitutes a small 
deviation from thermal equilibrium amenable to a simplified description. 
In this case, eq.~(\ref{Ttilde}) reads
\be
\tilde{T}_K = \frac{T_K}{\Omega} + \, ... 
\ee
where the ellipsis denote small corrections which can be controlled in the 
above-mentioned adiabatic approximation. The need for such an adiabatic 
approximation to make sense of Hawking radiation and the thermodynamics of 
time-evolving horizons has been stressed clearly as a requisite for the 
applicability of the Hamilton-Jacobi method in the tunneling approach
\cite{DiCriscienzoNadaliniVanzoZerbiniZoccatelli07} to 
the computation of $T$ \cite{AlexFirouzjaee} and in the renormalization of 
the scalar field stress-energy tensor in the background of cosmological 
black holes \cite{SaidaHaradaMaeda}.

\subsection{Locating the tilded apparent horizons}

The apparent horizons of a spherically symmetric spacetime (\ref{3bis}) 
are given by the roots of the equation $\nabla^cR\nabla_c R=0$, 
where $R$ is its areal radius, 
which is equivalent to $g^{RR}=0$ ({\em e.g.}, \cite{NielsenVisser06}). 
In the tilded world (\ref{3ter}) the 
tilded apparent horizons are given by $\tilde{g}^{\tilde{R}\tilde{R}}=0$ 
or, according to eq.~(\ref{19})
\be 
\frac{ 
A \left( \Omega_{,R}R+\Omega \right)^2 -\Omega_{,t}^2 B R^2 }{
\Omega^2 AB }  =0 
\ee
if $ \Omega_{,R}R+\Omega  \neq 0$. 
The possible solutions of this equation include the solutions of 
\be\label{27}
\frac{1}{B}=0 
\ee
(corresponding to $g^{RR}=0$ in the non-tilded spacetime (\ref{3bis})); 
those of 
\be\label{28}
A \left( \Omega_{,R}R+\Omega \right)^2 = \Omega_{,t}^2 B R^2 \,;
\ee
and $A=\infty$. Let us now look at some examples to make sense of the formulas 
derived so far.

\subsection{Example: FLRW space}

Let us consider again the spatially flat FLRW space conformal to 
Minkowski space as an example. We have here $A=B=1$, $F=1$, 
$\Omega=a(\eta)$, $R=r$, $\tilde{R}=ar$, $K^a= \left( \partial/\partial 
\eta \right)^a$, and
\be \label{29}
\tilde{K}^0=\frac{a K^0}{ a \tilde{A} \sqrt{
1+ \frac{a_{\, \eta} a^2 r^2}{a^2\left[ a^2 -\frac{a_{,\eta}^2 a^2 r^2}{a^2}
\right]}}} \,.
\ee
The relation  $a_{,\eta}=a\dot{a}$ with $H \equiv \dot{a}/a$ gives
\be\label{25}
\tilde{K}^{\mu} =\left( \frac{K^0}{a^2 \sqrt{1-H^2 \tilde{R}^2} }, \vec{0} \right) 
\ee
and the norm squared 
\be
\tilde{K}^a\tilde{K}_a= -\frac{1}{a^2}  \,.
\ee
The Kodama vector of FLRW space is timelike ($\tilde{K}^a\tilde{K}_a<0$) 
in the region below the cosmological horizon $\tilde{R}<1/H$ in which it is defined.

\subsection{Example: the Sultana-Dyer spacetime}

Let us return to the example of the Sultana-Dyer black hole and let us 
locate the apparent horizons using eqs. (\ref{27})-(\ref{28}). For 
comparison, we refer to the study of the causal structure of this 
spacetime in Ref.~\cite{SaidaHaradaMaeda}. The apparent horizons are the 
roots of the equation 
$ 1-\frac{2\tilde{M}_{MSH}}{\tilde{R}}=\tilde{g}^{ab} \tilde{\nabla}_a 
\tilde{R} \tilde{\nabla}_b \tilde{R}=\tilde{g}^{\tilde{R}\tilde{R}}=0$ . 
Since $\tilde{R}=ar$, the expression (\ref{8plus4}) of the 
Misner-Sharp-Hernandez mass $\tilde{M}_{MSH}$ in the Sultana-Dyer metric 
gives 
\be 
2ma+\frac{r^3 a_{,\tau}^2}{a} \left( 1+\frac{2m}{r} 
\right)-4mra_{,\tau}=ar \,. 
\ee 
Using the fact that $a(\tau)=\tau^2 
$ and $a_{,\tau}=2\tau$, one obtains \be 1-\frac{2m}{r}= 
\frac{4r^2}{\tau^2} \left( 1+\frac{2m}{r} \right) -\frac{8m}{\tau} 
\ee 
as the equation locating the apparent horizons (which coincides with 
eq.~(3.10) of \cite{SaidaHaradaMaeda}). This is the 
cubic equation for 
$r$ 
\be \label{cubic}
4r^3 +8mr^2 -\left( 8m+\tau \right)\tau r +2m\tau^2 =0 \,,
\ee
the real positive roots of which are \cite{SaidaHaradaMaeda}
\begin{eqnarray}
r_1 &=& \frac{-4m -\tau +\sqrt{ \tau^2 +24m\tau +16m^2}}{4} \,,\\
&& \nonumber\\
r_2 &=& \frac{\tau ( \eta,r)}{2} \,,
\end{eqnarray} 
with $r_1<r_2$. In addition, the surface $r=2m$, the null event horizon 
of the Schwarzschild seed metric, remains an event horizon for the 
Sultana-Dyer metric (see Ref.~\cite{SaidaHaradaMaeda} for a conformal 
diagram of this spacetime).

Let us proceed now using eqs.~(\ref{27})-(\ref{28}). Eq.~(\ref{27}) 
gives again $r_{EH}=2m$, corresponding to areal radius $\tilde{R}_{EH}
=ar_{EH}=2m 
\tau^2$ and to an event horizon which is exactly comoving with the 
cosmic substratum (this fact, not noted before, has some relevance for 
the long-standing debate of the effect of the cosmological expansion on 
local systems, see Ref.~\cite{CarreraGiuliniRMD10} for a 
recent review). Eq.~(\ref{28}) gives 
\be 
\left( 1-\frac{2m}{r} \right) \left( 
a_{\,\tau}\tau_{,r}r+a \right)^2 =\frac{\left( 
a_{,\tau}\tau_{,t}\right)^2 r^2}{1-\frac{2m}{r}} 
\ee 
and
\be 
\left( 1-\frac{2m}{r} \right) \left( 
\frac{ 2m a_{,\tau}}{1-\frac{2m}{r}} +a \right)
=\pm  a_{,\tau} r  \label{plusminus} \,.
\ee 
The upper (positive) sign applies if $r>2m$. Strictly speaking, 
the region $r\leq 2m$ is not allowed because we are using here 
the conformal transformation (\ref{8plus1}) with $a=\tau^2 $ 
and 
$\tau$ given by eq.~(\ref{8plus3}), which requires $r>2m$. 
However, 
one can consider the continuation of the metric (\ref{8plus1}) 
to 
the region $r\leq 2m$, hence we keep also the lower (negative) 
sign in eq.~(\ref{plusminus}) for the moment.
 Then we obtain 
\be \label{quadratic} 
\epsilon \, a_{, \tau} r^2 +\left( 2ma_{, \tau} +a \right) 
r-2ma=0
\ee
and
\be
r=
\frac{ -\left( 4m+\tau \right)\pm \sqrt{ \left( 4m+\tau 
\right)^2 +16m\tau}}{4\epsilon} \,,
\ee
where $\epsilon =\pm 1 $ keeps track of both signs coming from 
eq.~(\ref{plusminus}). For $\epsilon= -1$ we obtain the roots
\be
r_1= \frac{\tau}{2} \,, \;\;\;\;\;\;\;\;
r_2 = 2m \,,
\ee
while for $\epsilon=+1$ one obtains
\be
r_3= -\left[ 
\frac{4m + \tau + \sqrt{ \tau^2 + 
24 m\tau + 16m^2 } }{4} \right] \,, 
\ee 
which is negative and is discarded, and 
\be
r_4= 
\frac{-\left( 4m + \tau \right) 
+ \sqrt{ \tau^2 + 24 m\tau + 16m^2 } }{4}  \,, 
\ee 
which is the radius of the black hole apparent horizon found by 
Saida, Harada, and Maeda (eq.~(3.11a) of 
\cite{SaidaHaradaMaeda}). They do not use the form of 
the Sultana-Dyer metric explicitly conformal to Schwarzschild 
and they do not need to invoke any continuation, although they 
arrive to the cubic equation (\ref{cubic}) for the apparent 
horizon radii which 
is less straightforward to solve than the quadratic 
equation~(\ref{quadratic}). 
In terms of the areal radius $R=ar$, the 
apparent horizon radii are  
\be
R_1=\frac{\tau^3}{2} \,, \;\;\;\;\;\;\;
R_2=R_{EH}=2m \tau^2 \,,
\ee
and  
\be
R_{4}=\frac{-4 m - \tau + \sqrt{ \tau^2 + 24 m\tau 
+ 16m^2 } }{4} \, \tau^2 \,.
\ee 
Ultimately these are implicit equations for the radii of the 
apparent horizons in terms of tilded quantities $t$ and 
$\tilde{R}$.

The Hawking temperature of the Sultana-Dyer black hole was studied in 
Ref.~\cite{SaidaHaradaMaeda} by computing the renormalized stress-energy 
tensor of a free, conformally coupled, scalar field in this spacetime. 
The result is the effective temperature
\be
\tilde{T}= \frac{1}{8\pi m a} + \, ...
\ee
where the ellipsis denotes corrections which are small in the limit of a 
slowly evolving black hole \cite{SaidaHaradaMaeda}. Since 
$T=\frac{1}{8\pi m} $ is the Hawking temperature of the ``seed''  
Schwarzschild black hole, one can write 
\be
\tilde{T}=\frac{T}{\Omega} + \, ... \,;
\ee
The conformal factor of the Sultana-Dyer black hole does not depend on 
the radial coordinate and, in the adiabatic approximation in which its 
time variation is small, the Hawking temperature does have the scaling 
behaviour expected on dimensional grounds. This scaling law, however, will 
break down as soon as the conformal transformation is allowed to be 
radial-dependent, or the apparent horizon is allowed to vary rapidly.

\subsection{Asymptotically flat spacetimes}

The transformation properties of the Hawking temperature of black hole 
horizons under conformal transformations have been the subject of much 
literature, in which asymptotically flat spacetimes were considered 
together with conformal transformations which reduce to the identity 
($\Omega \rightarrow 1$) at spatial infinity $R\rightarrow +\infty$ 
\cite{JacobsonKang, JacobsonKangMyers}. 
This last property follows from the necessity of normalizing to 
unity the Killing vector at spatial infinity, where observers detecting 
the Hawking flux are located. It is clear that this case is not relevant 
when the interest is on  cosmological black holes, however it has its importance 
and it is worth at least a brief mention here. In this special case the transformation 
property~(\ref{7}) of the Misner-Sharp-Hernandez mass gives 
$\tilde{M}_{MSH} =M_{MSH}$ at infinity, {\em i.e.}, this mass notion is a 
conformal invariant when evaluated at large spatial distances from the 
asymptotically flat inhomogeneity. 

Similarly, eq.~(24) relating the Kodama vectors in the tilded and 
non-tilded worlds reduces, in this special case, to $\tilde{K}^a \simeq 
K^a $ at infinity, or wherever one wants to normalize the timelike Killing 
vector to unity. This fact is important because the Kodama vector defines a 
specific (``Kodama'') surface gravity and a Kodama temperature which is 
regarded by many authors as the physical temperature of a black hole. The 
invariance of the Kodama vector under restricted conformal transformations 
therefore implies the conformal invariance of the temperature.

\section{\label{K-S}Kerr-Schild transformations}

Another technique employed to generate cosmological black holes embedded 
in FLRW ``backgrounds'' is that of (generalized) Kerr-Schild transformations 
\cite{KerrSchild}: beginning from the ``seed'' metric $g_{ab}$, this 
transformation generates
\be
g_{ab} \rightarrow \bar{g}_{ab}=g_{ab}+2\lambda \, l_al_b \,,
\ee
where $\lambda $ is a function of the spacetime point and $l^a $ is a 
null and geodesic vector of the metric $g_{ab}$,
\be
g_{ab}l^a l^b=0 \,, \;\;\;\;\;\;  
l^a \nabla_a l^b=0 \,.
\ee
It follows that $l^a $ is null and geodesic also with respect to the 
``new'' metric $\bar{g}_{ab}$,
\begin{eqnarray}
\bar{g}_{ab} l^a l^b &=& g_{ab}l^a l^b +2\lambda l_a l^a l_b l^b =0 \,, \\
&& \nonumber\\ 
\bar{g}_{ab} l^a \nabla^b l^c &=& 0 \,.
\end{eqnarray}
The tensor  
\be\label{inverseKS} 
\bar{g}^{ab}= g^{ab}-2\lambda l^al^b 
\ee
 is the inverse of the 
metric $\bar{g}_{ab}$, as follows from the fact that 
\begin{eqnarray}
 \bar{g}^{ab} \bar{g}_{bc} &=& 
\left( g^{ab} -2\lambda l^al^b \right) 
\left( g_{bc} +2\lambda l_b l_c \right)\nonumber\\ 
&&\nonumber\\
&=& 
g^{ab}g_{bc} +2\lambda g^{ab} l_b l_c 
-2\lambda g_{bc} l^al^b -4\lambda^2 l^a l^b l_b l_c \nonumber\\
&&\nonumber\\
&= &  \delta^a_c \,. 
\end{eqnarray}
The {\em caveat} raised for metrics generated by conformal 
transformations applies here: it is not trivial that a seed metric which 
solves the Einstein equations will generate another solution of the same 
equations corresponding to a ``reasonable'' form of mass-energy, and 
most of the cosmological black hole solutions generated by this 
technique in the literature have unphysical stress-energy tensors 
at least in some regions of the spacetime 
manifold \cite{KerrSchild}. 
However, this technique is not to be discarded, for example it 
generates the Schwarzschild and Reissner-Nordstr\"om solutions using 
Minkowski space as the seed. 

We can ask ourselves how the Misner-Sharp-Hernandez mass transforms 
under Kerr-Schild transformations which use a spherically symmetric 
metric of the form (\ref{3bis}) and respects the spherical symmetry, {\em  
i.e.}, $\lambda=\lambda (t, R)$ and $l^a=l^a (t, R) $ in these 
coordinates. We have
\begin{eqnarray}
d\bar{s}^2 &=& ds^2 +2\lambda l_al_b dx^a dx^b =
-\left[ A- 2\lambda (l_0)^2 \right] dt^2 \nonumber\\
&&\nonumber\\
&\, & +\left[ B+2\lambda 
(l_1)^2 \right] dR^2 +4\lambda l_0 l_1 dtdR +R^2 d\Omega_{(2)}^2 \,,\nonumber\\
&& 
\end{eqnarray}
where the ``new'' areal radius coincides with the ``old'' one, 
$\bar{R}=R$. In order to eliminate the $dtdR$ cross-term we introduce 
again a new time coordinate $T$ defined by $dT=\frac{1}{F} \left( 
dt+\beta dR \right)$, where $\beta (t, R)$ is a function to be 
determined and $F(t, R)$ is an integrating factor satisfying 
eq.~(\ref{eqforF}). The relation $dt=FdT-\beta dR$ gives
\begin{eqnarray}
d\bar{s}^2 &=&  
-\left[ A-2\lambda (l_0)^2 \right] F^2 dT^2 \nonumber\\
&&\nonumber\\
&\, & +
\left\{ B+2\lambda (l_1)^2 -\beta^2 \left[ A-2\lambda (l_0)^2 \right] 
-4\lambda l_0 l_1 \beta \right\}  dR^2 \nonumber\\
&&\nonumber\\
&\, &  +2F \left\{ 
\beta \left[ A-2\lambda (l_0)^2 \right] +2\lambda l_0 l_1 \right\} dTdR
+R^2 d\Omega_{(2)}^2  \,.\nonumber\\
&&\nonumber
\end{eqnarray}
By imposing that
\be
\beta (t, R)= \frac{-2 \lambda l_0 l_1}{A-2\lambda (l_0)^2}
\ee
the metric is diagonalized and assumes the form
\begin{eqnarray}
d\bar{s}^2 &=& 
-\left[ A-2\lambda (l_0)^2 \right] F^2 dT^2 \nonumber\\
&&\nonumber\\
&\, & +
\left\{ B+ 2\lambda (l_1)^2 
+ \frac{ 4\lambda^2 (l_0)^2 (l_1)^2 }{ 
A-2\lambda (l_0)^2 } \right\}  dR^2 
+R^2 d\Omega_{(2)}^2  \,.\nonumber\\
&& 
\end{eqnarray}
Given that $\bar{R}=R$, the Misner-Sharp-Hernandez mass of the 
barred spacetime is defined by
\be
1-\frac{2\bar{M}_{MSH}}{R} =\bar{g}^{ab} \nabla_a R \nabla_b R =
\bar{g}^{RR}
\ee
which gives, using eq.~(\ref{inverseKS}),
\be \label{MSHmassKS}
\bar{M}_{MSH}= M_{MSH} +\lambda (l^1)^2 R \,.
\ee
The ``new'' Misner-Sharp-Hernandez mass of  a sphere of radius $R$ 
is always non-negative  if $M_{MSH}$ is non-negative. 

It is straightforward to locate the apparent horizons of the spherical 
metric $\bar{g}_{ab}$ in terms of those of the metric $g_{ab}$. These 
apparent horizons are the roots of $\bar{g}^{ab} \bar{\nabla}_a R 
\bar{\nabla}_b R =0$, or 
$\bar{g}^{RR}=0$, which gives
\be\label{KSAHs}
g^{RR}-2\lambda (l^1)^2 =0 \,.
\ee
In practice, a null future-oriented vector is defined up to a positive 
constant and one can set $l^1=1$ in eqs.~(\ref{MSHmassKS}) and 
(\ref{KSAHs}) without loss of generality. It is clear then that it is the 
function $\lambda$ which determines the deviation of the barred quantities 
from the non-barred ones.

\subsection{Behavior of the Kodama temperature under Kerr-Schild 
transformations}

The behavior of the Kodama temperature under Kerr-Schild transformations 
preserving the spherical symmetry of the metric is derived using 
eq.~(\ref{surfacegravity}) for the Kodama surface gravity. By using the 
fact that $\tilde{h}_{ij}=h_{ij} +2\lambda l_il_j$ and
$\tilde{h}^{ij}=h^{ij} -2\lambda l^i l^j$, one finds
\begin{eqnarray}
\bar{h} &=& \left[ -A+2\lambda (l_0)^2\right] 
\left[ B+2\lambda (l_1)^2\right]  -4\lambda^2 (l_0)^2 (l_1)^2 \nonumber\\
&&\nonumber\\
&=& 
-AB +2\lambda \left[ -A(l_1)^2 +B(l_0)^2 \right] \nonumber\\
&&\nonumber\\
&=& h\left\{ 1-2\lambda \left[ 
-\frac{(l_1)^2}{B} +\frac{(l_0)^2}{A} \right]\right\}
\,,
\end{eqnarray}
where $h=-AB$. Remembering that $l^a$ is a null vector of the metric 
$g_{ab}$, one obtains $-A(l^0)^2+B(l^1)^2=0$ and, using $l_0=g_{00}l^0=-Al^0$, 
$ l_1=g_{11} l^1 = Bl^1$, it is $-\frac{(l_0)^2}{A}+ \frac{(l_1)^2}{B}=0$ and
the exact relation 
\be
\bar{h}=h
\ee
follows.  Eq.~(\ref{surfacegravity}) now yields 
\begin{widetext} 
\begin{eqnarray}
\bar{\kappa}_K &=& 
\frac{1}{2\sqrt{-\bar{h}}} \left\{ 
\partial_t \left[ \sqrt{-\bar{h}} \left( 
\bar{h}^{00}\partial_t R +\bar{h}^{01}\partial_R R \right)\right] 
+ \partial_R \left[ \sqrt{-\bar{h}} \left( 
\bar{h}^{11}\partial_R R +\bar{h}^{10}\partial_t  R \right)\right] \right\}\nonumber\\
&&\nonumber\\
&=& \frac{1}{2\sqrt{-h}} \partial_R \left( \frac{ \sqrt{-h}}{B}\right) 
-\frac{1}{\sqrt{AB}} \left[ 
\partial_t \left( \lambda l^0 l^1 \sqrt{AB} \right)
+ \partial_R \left( \lambda l^1 l^1 \sqrt{AB} \right) \right] \,,
\end{eqnarray}
or
\be \label{KSsurfacegravity}
\bar{\kappa}_K = \kappa_K   
-\frac{1}{\sqrt{AB}} \left[ 
\partial_t \left( \lambda l^0 l^1 \sqrt{AB} \right)
+ \partial_R \left( \lambda l^1 l^1 \sqrt{AB} \right) \right] \,.
\ee
Again, one can set $l^1=1$ in the previous expressions.
\end{widetext}

\subsection{Example: the Reissner-Nordstr\"om black hole}
 
The Reissner-Nordstr\"om black hole with line element
\be\label{RN}
d\bar{s}^2 =- \left( 1-\frac{2m}{r}+\frac{Q^2}{r^2} \right) dT^2 +
\frac{dr^2}{1-\frac{2m}{r}+\frac{Q^2}{r^2} } +r^2 d\Omega_{(2)}^2 \,,
\ee
where $m$ and $Q$ are the mass and charge parameters, 
respectively, can be obtained by a Kerr-Schild transformation of the 
Minkowski space metric $ds^2=-dt^2+dr^2 +r^2d\Omega_{(2)}^2$ in 
spherical coordinates with
\begin{eqnarray}
\lambda (t, r) &=& \frac{m}{r} -\frac{Q^2}{2r^2} \,, \label{lambdaRN} \\
&&\nonumber\\
l^{\mu} &=& \left( -1,1, 0, 0 \right) \label{lRN}
\end{eqnarray} 
in these coordinates. In fact, this transformation gives
\begin{eqnarray}
d\bar{s}^2 &=& - \left( 1-\frac{2m}{r}+\frac{Q^2}{r^2} \right) dt^2 +
\left( 1+\frac{2m}{r}-\frac{Q^2}{r^2} \right) dr^2 \nonumber\\
&&\nonumber\\
&\, & +r^2 d\Omega_{(2)}^2 +2\left( \frac{2m}{r}-\frac{Q^2}{r^2} \right) 
dtdr \,.
\end{eqnarray}
This line element can be brought to the diagonal form (\ref{RN}) by 
performing the usual transformation to a new time coordinate $T$ defined 
by $dT=\frac{1}{F} \left( dt +\beta dr\right)$, which yields 
\begin{eqnarray}
d\bar{s}^2 &=& - \left( 1-\frac{2m}{r}+\frac{Q^2}{r^2} \right) F^2 dT^2
\nonumber\\
&\, &  +
\left\{ -\left( 1-\frac{2m}{r}+\frac{Q^2}{r^2} \right)\beta^2 
+ 1 +\frac{2m}{r}-\frac{Q^2}{r^2} \right.\nonumber\\
&&\nonumber\\
&\, & \left.  
-2\left( \frac{2m}{r}-\frac{Q^2}{r^2}\right) \beta \right\} dr^2 \nonumber\\
&&\nonumber\\
&\, & 
+2 \left\{ \beta \left(1- \frac{2m}{r}+ \frac{Q^2}{r^2} \right) 
+\left(  \frac{2m}{r} - \frac{Q^2}{r^2} \right) \right\} F  
dTdr  \nonumber\\
&&\nonumber\\
&\, &  +r^2 d\Omega_{(2)}^2 \,.
\end{eqnarray}
The choice of the function
\be
\beta(t,r)= \frac{- \left( \frac{2m}{r}- \frac{Q^2}{r^2} \right)}{
1-\frac{2m}{r}+ \frac{Q^2}{r^2}   }
\ee
brings the line element into the form 
\begin{eqnarray}
d\bar{s}^2 &=& - \left( 1-\frac{2m}{r}+\frac{Q^2}{r^2} \right) F^2 dT^2 \nonumber\\
&&\nonumber\\
&\, &  +
\frac{dr^2}{1-\frac{2m}{r}+\frac{Q^2}{r^2} } +r^2 d\Omega_{(2)}^2 \,.\label{81}
\end{eqnarray}
The integrating factor $F(t,r)$ must satisfy eq.~(\ref{eqforF}), which 
for the static case under consideration reduces to
\be
\frac{\partial}{\partial r}\left( \frac{1}{F} \right) =0
\ee
and is satisfied by the choice $F \equiv 1$, which finally 
brings the metric~(\ref{81})  
into the familiar Reissner-Nordstr\"om form~(\ref{RN}). 

Our formula~(\ref{MSHmassKS}) for the Kerr-Schild transformation of the 
Misner-Sharp-Hernandez mass gives, using eqs.~(\ref{lambdaRN}), 
(\ref{lRN}), and $M_{MSH}=0$, the mass of a sphere of radius $r$ as 
\be 
\bar{M}_{MSH}=m-\frac{Q^2}{2r} \,. 
\ee 
This is the well known expression of the Misner-Sharp-Hernandez mass of 
a sphere of radius $r$ in the Reissner-Nordstr\"om spacetime, which can 
be obtained immediately from the definition $ 1-\frac{2\bar{M}_{MSH}}{R} 
= \bar{g}^{ab} \bar{\nabla}_a R \bar{\nabla}_b R$.

The apparent horizons of the metric $\bar{g}_{ab}$ can be obtained from 
those of the seed Minkowski metric (which has no apparent horizons) by 
using eq.~(\ref{KSAHs}) which yields, using eqs.~(\ref{lambdaRN}) and 
(\ref{lRN}), 
\be
g^{RR}-2\lambda (l^1)^2= 1-\left( \frac{2m}{r} -\frac{Q^2}{r^2} \right)=0
\ee
or $ r^2-2mr+Q^2=0$, which has the well known roots
\be
r_{\pm}= m \pm \sqrt{m^2-Q^2} 
\ee
corresponding to the outer (event) horizon and to the inner (Cauchy) 
horizon, respectively. 

Finally, let us check that the Reissner-Nordstr\"om surface gravity 
coincides with that given by eq.~(\ref{KSsurfacegravity}). The latter 
equation gives, using $\kappa_K=0$ for Minkowski space, $AB=1$ and 
$\lambda=\frac{m}{r}-\frac{Q^2}{2r^2}$,
\begin{eqnarray}
\bar{\kappa}_K &=& 0-  \left\{
\partial_t \left[ \left( \frac{m}{r} -\frac{Q^2}{2r^2} \right)\right] 
+ \partial_r \left[ \left( \frac{m}{r} -\frac{Q^2}{2r^2} \right)\right]
\right\} \nonumber\\
&&\nonumber\\
&=&\frac{m^2}{r^2}-\frac{Q^2}{r^3} \,. \label{castoro}
\end{eqnarray} 
Direct use of the expression (\ref{surfacegravity}) gives
\begin{eqnarray}
\bar{\kappa}_K &=& \frac{1}{2} \left[ \partial_t \bar{h}^{00} + \partial_r 
\bar{h}^{11}\right] =\frac{1}{2} \partial_r \left( 
1-\frac{2m}{r}+\frac{Q^2}{r^2} \right) \nonumber\\
&&\nonumber\\
&=& \frac{m^2}{r^2}-\frac{Q^2}{r^3} \,,
\end{eqnarray}
in agreement with the previous expression~(\ref{castoro}).

\section{\label{Discussion}Discussion and conclusions}

The textbook concept of event horizon is suitable for stationary black 
holes and is central in the construction of black hole mechanics and 
thermodynamics but, due to its teleological nature, becomes useless for 
dynamical black holes where its best replacement currently 
available is the notion of apparent or trapping horizon. Here we restricted 
ourselves to spherical symmetry, for which the shortcoming of dependence 
of the apparent horizons on the spacetime foliation is less severe.

Conformal transformation techniques and Kerr-Schild transformations are 
commonly encountered in relation with dynamical black holes, in both 
General Relativity and alternative theories of gravity. When trying to 
extend the known theory of thermodynamics to apparent or trapping 
horizons in time-dependent situations \cite{AHthermodynamics}, it is 
useful to know how various quantities (Misner-Sharp-Hernandez mass or 
quasi local Hawking-Hayward energy, Kodama vector, surface gravity and 
temperature) transform under these transformations. We have addressed 
this question here: these quantities do not transform as one would expect 
on the basis of naive dimensional arguments \cite{Dicke}, which is 
natural since, for example, the quasi-local mass of a sphere depends on 
both the matter in it and the gravitational energy, while the naive 
arguments are designed for test particles. We have provided examples to test 
the transformation formulas that we derived in more intuitive situations.

In spite of rather strong claims, it is not yet clear whether marginally 
trapped surfaces and apparent/trapping horizons are the best 
concepts to apply in the characterization of black holes but, {\em de 
facto}, they are used routinely in numerical relativity with the 
practical goal of producing templates for immediate use in the 
detection of gravitational waves by giant laser interferometers such as 
{\em LIGO} and {\em VIRGO}. It is hoped that theoretical efforts 
be multiplied to understand better key concepts such as the practical 
definition of black hole, horizons, their theoretical thermodynamics, 
and the limitations of the theoretical constructs used.

\begin{acknowledgments}

This work is supported by Bishop's University and by the Natural 
Sciences and Engineering Research Council of Canada. VV is 
supported by FCT - Portugal through the grant SFRH/BPD/77678/2011. 
VF is grateful for the hospitality at CENTRA, where this work 
was begun.

\end{acknowledgments}


\end{document}